\documentstyle[12pt]{article}
\begin{document}
\def\l{\lambda}
\vskip 0.5in
\begin{center}
{\Large{\bf Topological Solitons in Chern-Simons Theories for Double-Layer
Fractional Quantum Hall Effect  }}
\vskip 0.7in
{\Large Ikuo Ichinose\footnote{e-mail address:
ikuo@hep1.c.u-tokyo.ac.jp}  and Akira Sekiguchi}

\vskip 0.2in
Institute of Physics, University of Tokyo, Komaba, Tokyo, 153 Japan\\
\end{center}
\vskip 0.2in
\begin{center} 
\begin{bf}
Abstract
\vskip 0.5in
\end{bf}
\end{center} 
Topological excitations in Chern-Simons (CS) gauge theories which describe double-layer
fractional quantum Hall effect are studied.
We shall consider generic $(m,m,n)$ Halperin state.
There are two type of solitons; one is vortex type excitation which has essentially the
same structure with quasi-hole excitation in the single-layer case.
The other is nontrivial pseudospin texture which is so-called skyrmion or meron.
We shall first study qualitative properties of solitons in the original CS gauge theory  
and give results of numerical calculation.
Then by using duality transformation, we derive an effective theory for topological
excitations in the fractional quantum Hall effect. 
For spin texture, that theory is nonrelativistic CP$^1$ nonlinear $\sigma$-model with a CS gauge
interaction and a Hopf term.
Finally, we study quantum mechanical system of multi-soliton states, retaining only center
coordinates of solitons as collective coordinates.
Existence of inter-layer tunneling drasically changes excitation spectrum.

\newpage

\section{Introduction}
\setcounter{footnote}{0}
Study of $2$-dimensional electron system in a strong magnetic field is one of
the most interesting topics in the condensed matter physics.
Especially in the last few years, both experimentally and theoretically, existence of
topologically stable excitations is suggested for quantum Hall state (QHS) with internal
degrees of freedom like spin singlet QHS and QHS in double-layer systems.
Theoretical study started with QHS at filling factor $\nu=1$ by taking into account
spin degrees of freedom\cite{sondhi}.
It was suggested that low-energy charged excitation in that system is {\em not}
a single-body electron but a topological spin texture with finite size and that 
excitation has similar structure with skyrmion in ferromagnet.
At present, some experiments support this observation\cite{exp}.

Somewhat detailed study on the double-layer case is given in Ref.\cite{moon}
by using lowest Landau
level (LLL) projection in the first-quantization language.
Most of studies on double-layer QHS given so far are restricted to Halperin's 
$(m,m,m)$ states.

In this paper, we shall study topological excitations in generic $(m,m,n)$ states
in the framework of the Chern-Simons (CS) gauge theory.
In the CS gauge theory for QH effect, the original fermionic electron is expressed 
by bosonized electron with odd number of CS flux quanta attached\cite{CS}.
QHS is characterized by the coherent condensation of bosonized electron.
In the single layer case, there is a topological excitation, which has a close resemblance to
the vortex in the superfluid, and that excitation is identified with Laughlin's quasi-hole excitation
in the fractional quantum Hall (FQH) state\cite{Laughlin}.
In the QHS with an internal symmetry, other topological excitations are expected
which have nontrivial structure in the internal space.

This paper is organized as follows.
In Sec.2, we shall explain models, which describe FQH effect (FQHE) in double layer (DL)
electron systems.
In general, there are two CS gauge fields corresponding to $U(1)\otimes U(1)$ local
symmetry, and two bosonized electron fields, which represent electrons in upper and lower
layers, respectively.
Ground state at specific filling factor has coherent condensation of both bosonized 
electron fields.
In Sec.3, it is explained that there are essentially two types of topological solitons
in the models. 
One is vortex which has nontrivial winding number for $U(1)$ symmetry corresponding
to the simultaneous phase rotation of the pseudospin up and down bosonized electron fields.
The other is meron which has nontrivial topological structure in the $SU(2)$ 
internal space of pseudospin.
Qualitative study of the topological solitons and numerical solutions
to field equations are given.
In Sec.4, an effective field theory for topological excitations is derived by
using duality transformation.
We first decompose bosonized electron fields into amplitude or density, common $U(1)$
phase factor and $CP^1$ variable.
Lowest Landau level condition is obtained explicitly, which determines deviation
of charge density in terms of topological currents of vortex and meron.
Pseudospin $CP^1$ part of the effective field theory is a $CP^1$ nonlinear $\sigma$-model
interacting with a CS gauge field and with a Hopf term.
In Sec.5, quantum mechanics of solitons is discussed.
Only center coordinates of them are taken into account as collective coordinates.
Effect of inter-layer tunneling is considered.
Single meron cannot exist with finite energy.
Spectrum of meron pair is studied.
Section 6 is devoted for discussion and conclusion.

%%%%%%%%%%%%%%%%%%%%%%%%%%%%%%%%%%%%%%%%%%%%%%%%%%

\section{Models}
\setcounter{equation}{0}

Lagrangian of the model is given as follows,
\begin{eqnarray}
{\cal L}_{\psi}&=&-\bar{\psi}_{\uparrow}(\partial_0-ia^+_0-ia^-_0)\psi_{\uparrow}  
-\bar{\psi}_{\downarrow}(\partial_0-ia^+_0+ia^-_0)\psi_{\downarrow}   \nonumber   \\
&&\; -{1 \over 2M}\sum_{\sigma=\uparrow,\downarrow}
|D^{\sigma}_j\psi_{\sigma}|^2-V[\psi_{\sigma}],  
\label{Lpsi}   \\
{\cal L}_{\mbox{\small CS}}&=&{\cal L}_{\mbox{\small CS}}(a^+_{\mu})+
{\cal L}_{\mbox{\small CS}}(a^-_{\mu})   \nonumber   \\
&=&-{i\over 4}\epsilon_{\mu\nu\lambda}\Big({1\over p}a^+_{\mu}
\partial_{\nu}a^+_{\lambda}
+{1\over q}a^-_{\mu}\partial_{\nu}a^-_{\lambda}\Big),
\label{LCS}
\end{eqnarray}
where $\psi_{\sigma}$ ($\sigma=1,2$ or $\uparrow,\downarrow$) is the bosonized 
electron fields in upper and lower layers,
respectively, $M$ is the mass of electrons and $p$ and $q$ are parameters.
Greek indices take $0,1$ and $2$, while roman indices take $1$ and $2$.
$\epsilon_{\mu\nu\lambda}$ is the antisymmetric tensor and the
covariant derivative is defined as 
\begin{equation}
D^{\uparrow\downarrow}_j=\partial_j-ia^+_j\mp ia^-_j+ieA_j,
\end{equation}
where external magnetic field is directed to the $z$-axis and in the symmetric Coulomb gauge
$A_j=-{B \over 2}\epsilon_{jk}x_k$.
$V[\psi_{\sigma}]$ is interaction term between electrons like the Coulomb repulsion
and short-range four-body interaction, e.g.,
\begin{equation}
g_1\Big(|\psi_{\uparrow}|^4+|\psi_{\downarrow}|^4\Big)+g_2|\psi_{\uparrow}
\psi_{\downarrow}|^2, \;\;  \mbox{etc.}
\label{interaction}
\end{equation}
Chern-Simons constraints are obtained by differentiating the Lagrangian
with respect to $a^{\pm}_0$
\begin{eqnarray}
\epsilon_{ij}\partial_ia^+_j&=&2p\bar{\Psi}\Psi  \nonumber   \\
\epsilon_{ij}\partial_ia^-_j&=&2q\bar{\Psi}\sigma_3\Psi  \nonumber   \\
\Psi&=&(\psi_{\uparrow} \;\; \psi_{\downarrow})^t,
\label{CScon}
\end{eqnarray}
where $\epsilon_{ij}=\epsilon_{0ij}$.
As we shall see, $(p+q)$ must be an odd integer times $\pi$ and $(p-q)$
must be an integer times $\pi$ for the original electrons to be fermionic.

Partition function in the imaginary-time formalism is given as 
\begin{equation}
Z=\int [D\bar{\psi}D\psi Da]\exp \{\int d\tau d^2x ({\cal L}_{\psi}+{\cal L}_{\mbox{\small CS}})\}.
\end{equation}
The Lagrangian ${\cal L}={\cal L}_{\psi}+{\cal L}_{\mbox{\small CS}}$ is invariant under 
$U(1)\otimes U(1)$ gauge transformation,
\begin{eqnarray}
\Psi &\rightarrow& e^{i\theta_1+i\theta_2\sigma_3}\Psi  \nonumber    \\
a^+_{\mu}  &\rightarrow&  a^+_{\mu}+\partial_{\mu}\theta_1   \label{gaugetrn}  \\
a^-_{\mu}  &\rightarrow&  a^-_{\mu}+\partial_{\mu}\theta_2.  \nonumber
\end{eqnarray}

Ground state for fractional quantum Hall effect is given by the following
field configuration,
\begin{eqnarray}
&&\psi_{\uparrow,0}=\sqrt{\bar{\rho}_{\uparrow}}=\sqrt{\bar{\rho}/2},  \nonumber  \\
&&\psi_{\downarrow,0}=\sqrt{\bar{\rho}_{\downarrow}}=\sqrt{\bar{\rho}/2},  \nonumber  \\
&&eB=\epsilon_{ij}\partial_ia^+_{j,0}=2p\bar{\rho}, \; \; \epsilon_{ij}\partial_ia^-_{j,0}=0, 
\;\; a^+_{j,0}=eA_j,  \label{groundcon}   \\
&& \nu={2\pi\bar{\rho} \over eB}={\pi \over p},
\label{ground}
\end{eqnarray}
where $\bar{\rho}$ is the average electron density.
It is easily verified that the above static configuration is a solution to the field equations
if and only if the filling factor $\nu$ has specific values given by (\ref{ground}).

One may think that there is a massless mode, because the configuration (\ref{groundcon})
breakes the $U(1)\otimes U(1)$ symmetry.
However, these massless modes are eaten by the CS gauge fields and the CS gauge
fields acquire mass term by the Higgs mechanism.
This is important, because QHS must be incompressible and there must be no
gapless modes.

It is not so difficult to show that the above ground state (\ref{groundcon}) corresponds
to the following Halperin's $(m,m,n)$ state\cite{halperin} 
in the first-quantization language\cite{ezawa};
\begin{eqnarray}
\Psi_G(z^{\uparrow};z^{\downarrow})&=&\prod_{i<j}(z^{\uparrow}_i-z^{\uparrow}_j)^m
\prod_{i<j}(z^{\downarrow}_i-z^{\downarrow}_j)^m
\prod_{i,j}(z^{\uparrow}_i-z^{\downarrow}_j)^n  \nonumber  \\
&&\; \times \exp \Big(-{eB\over 4}(\sum |z^{\uparrow}_i|^2+\sum |z^{\downarrow}_i|^2)\Big),
\label{halperin}
\end{eqnarray}
where $m={1\over \pi} (p+q)$ and $n={1\over \pi} (p-q)$.
For the specific case $m=n$, one can set $a^-_j=0$ and only the single CS gauge field
remains.

%%%%%%%%%%%%%%%%%%%%%%%%%%%%%%%%%%%%%%%%%%%%%%%%%

\section{Solitons and numerical calculations}
\setcounter{equation}{0}
In this section, we shall study topological solitons in the CS gauge theory 
(\ref{Lpsi}) and (\ref{LCS}).
We first assume the following specific form of the potential between bosonized
electrons;
\begin{equation}
V[\psi_{\sigma}]={p\over M}\Big(\bar{\Psi}\Psi\Big)^2+
{q\over M}\Big(\bar{\Psi}\sigma_3\Psi\Big)^2+W[\psi_{\sigma}],
\label{pot}
\end{equation}
where $W[\psi_{\sigma}]$ represents long-range inter-layer and intra-layer
Coulomb repulsions.
It is argued that the above short-range repulsion is required in order to retain the Pauli
exclusion principle in the bosonization of fermionic field\cite{ezawa1}.
In the present case, field equation to be solved becomes very simple with this term
as we shall see.
It is expected, however, that the existence of this term influences 
behavior of  solutions only near the center of them.

It is easily verified that the following modified Bogomol'nyi decomposition holds,
\begin{eqnarray}
\sum_{\sigma}|D^{\sigma}_j\psi_{\sigma}|^2&=&\sum_{\sigma}|(D^{\sigma}_1-
iD^{\sigma}_2)\psi_{\sigma}|^2+\sum_{\sigma}\epsilon_{jk}\partial_j[\bar{\psi}
iD^{\sigma}_k\psi_{\sigma}]    \nonumber   \\
&&\; -\; \epsilon_{ij}\partial_i(a^+_j-eA_j)\Big(\bar{\Psi}\Psi\Big)-\epsilon_{ij}
\partial_ia^-_j\Big(\bar{\Psi}\sigma_3\Psi\Big).
\label{Bog}
\end{eqnarray}
By using the CS constraints (\ref{CScon}) and (\ref{pot}), 
Hamiltonian (in the real-time formalism) is given as 
\begin{equation}
{\cal H}={1\over 2M}\sum_{\sigma}|(D^{\sigma}_1-iD^{\sigma}_2)\psi_{\sigma}|^2
+{\omega_c \over 2}\Big(\bar{\Psi}\Psi\Big)+W[\psi_{\sigma}],
\label{Hamiltonian}
\end{equation}
where $\omega_c=eB/M$.
If we treat the Coulomb interaction as a perturbation, the lowest-energy configurations
satisfy the following ``self-dual" equations,
\begin{equation}
(D^{\uparrow}_1-iD^{\uparrow}_2)\psi_{\uparrow}=0, \;
(D^{\downarrow}_1-iD^{\downarrow}_2)\psi_{\downarrow}=0.
\label{self}
\end{equation}
Actually, (\ref{self}) is nothing but the lowest Landau level condition, and there are
many solutions to (\ref{self}).
This degeneracy is solved by the Coulomb interaction, and the ground state is the uniform
configuration (\ref{groundcon}).

We are going to seek topological soliton solutions to the self-dual equation (\ref{self}).
In this section, we are interested in single soliton configurations.
Therefore, we parameterize the fields as
\begin{eqnarray}
\psi_{\uparrow}&=&\sqrt{\bar{\rho}}\exp \Big(u_1(r)+in_1\theta (x)\Big),  \nonumber  \\
\psi_{\downarrow}&=&\sqrt{\bar{\rho}}\exp \Big(u_2(r)+in_2\theta (x)\Big),
\label{para1}
\end{eqnarray}
where $r=|\vec{x}|$, $\theta(x)$ is the azimuthal function 
$\theta(x)=\arctan \Big({x_2\over x_1}\Big)$,
and $n_1$ and $n_2$ are integers which label topological solitons.
By substituting (\ref{para1}) into (\ref{self}), we have
\begin{eqnarray}
a^+_k+a^-_k-eA_k&=&n_1\partial_k\theta-\epsilon_{kl}\partial_lu_1, \nonumber  \\
a^+_k-a^-_k-eA_k&=&n_2\partial_k\theta-\epsilon_{kl}\partial_lu_2.
\label{fieldeq0}
\end{eqnarray}
By differentiating Eq.(\ref{fieldeq0}) and  using the CS constraint (\ref{CScon}),
we have the following Toda-type equations for $u_i(r)$ $(i=1,2)$,
\begin{equation}
 {d^2 u_i\over d \l^2}+{1\over \l}{d u_i\over d \l}
+n_i{\delta (\l) \over \l}+2-2\sum_j K_{ij}e^{2u_j}=0,
\label{fieldeq1}
\end{equation}
\begin{equation}
 \l=\sqrt{p\bar{\rho}}\cdot r={r \over \sqrt{2}l_0}, \; \; K={1\over p} \left(
                                         \begin{array}{cc}
					 p+q & p-q  \\
					 p-q & p+q
					 \end{array}
					 \right),  \nonumber 
\end{equation}
where $l_0$ is the magnetic length $l_0=1/\sqrt{eB}$.
It is obvious that $u_1(r)$ and $u_2(r)$ couple with each other nontrivially
unless $p=q$.

It is useful to rewrite (\ref{fieldeq1}) by introducing $v_i(\l) \equiv u_i(\l)+n_i\ln \l$
as follows;
\begin{equation}
\Big({d^2\over d\l^2}+{1\over \l}{d\over d\l}\Big)v_i+2-2\sum_jK_{ij}e^{2v_j}\l^{-2n_j}=0,
\label{fieldeq2}
\end{equation}
where we have used the identity
\begin{equation}
\Big({d^2\over d\l^2}+{1\over \l}{d\over d\l}\Big)\ln \l={\delta (\l) \over \l}. \nonumber
\end{equation}
In terms of $v_i(\l)$,
\begin{eqnarray}
\psi_{\uparrow}&=&\sqrt{\bar{\rho}}\exp \Big(v_1(\l)-n_1\ln \l+in_1\theta (x)\Big),  \nonumber  \\
\psi_{\downarrow}&=&\sqrt{\bar{\rho}}\exp \Big(v_2(\l)-n_2\ln \l+in_2\theta (x)\Big).
\label{para2}
\end{eqnarray}
From (\ref{para2}), it is obvious that the integers $n_i$ must be nonpositive,
$n_i=0,-1,-2,...$, for single-valuedness and regularity of solution at the center of soliton,
and $v_i(\l)$'s are regular functions.
On the other hand, boundary condition at the infinity is given as 
\begin{equation}
\lim_{\l\rightarrow \infty}e^{u_i(\l)}={1\over \sqrt{2}},
\label{bc}
\end{equation}
which guarantees that at sufficiently far from the soliton, amplitude of solution is equal
to that of the ground state.

Before going into detailed study of the numerical calculation, let us examine
qualitative properties of  solitons.
There are essentially two types of topological solitons.
The first one has a nontrivial winding number for $U(1)$ symmetry of simultaneous
phase rotation of the both up and down bosonized electron fields.
This soliton corresponds to the $(n_1,n_2)=(-1,-1)$ in (\ref{para1}), and has trivial
structure for the pseudospin internal space.
We shall call this soliton vortex.
As we shall show in the numerical calculation, behavior of this soliton is very similar
to the vortex in the single-layer system.
The other ones are pseudospin texture and correspond to the $(-1,0)$ and 
$(0,-1)$ in (\ref{para1}).
Functions $u_1(\l)$ and $u_2(\l)$ for $(-1,0)$ behave as
\begin{eqnarray}
&& u_1(\l) \sim \ln \l, \; u_2(\l) \sim \mbox{const.}, \;\; \mbox{for}\; \; \l \rightarrow 0, \nonumber  \\
&& u_1(\l)=u_2(\l) \sim -\ln \sqrt{2}, \;\; \mbox{for}\;\; \l \rightarrow \infty,
\label{meroncon}
\end{eqnarray}
and similarly for the $(0,-1)$.
It is useful to introduce ``pseudospin operator" $\vec{m}(x)$ as usual 
by\footnote{ $\vec{m}(x)$ is not invariant under transformation (\ref{gaugetrn}).  
But under regular gauge transformation, topological structure of $\vec{m}(x)$
is invariant.   
Gauge-invariant definition of the topological charge will be given in Sec.4.},
\begin{equation}
\vec{m}=\bar{\Psi}\vec{\sigma}\Psi.
\label{spin1}
\end{equation}
From (\ref{meroncon}), it is obvious for the $(-1,0)$ meron
\begin{eqnarray}
&& \vec{m}(x) \sim -\hat{z}, \; \;\mbox{as} \;\; \vec{x} \rightarrow 0,  \nonumber  \\
&& \vec{m}(x) \sim \cos \theta (x) \cdot \hat{x}+\sin \theta (x)\cdot \hat{y},
\;\; \mbox{as} \;\; |\vec{x}| \rightarrow \infty,
\label{meroncon2}
\end{eqnarray}
where $\hat{x}$, etc., are unit vectors in the $x$, $y$ and $z$ directions.
Picture of the above configuration of $\vec{m}(x)$ is given in Fig.1.
Similar picture is obtained for $(0,-1)$ soliton;
\begin{eqnarray}
&& \vec{m}(x) \sim \hat{z}, \; \;\mbox{as} \;\; \vec{x} \rightarrow 0,  \nonumber  \\
&& \vec{m}(x) \sim \cos \theta (x) \cdot \hat{x}-\sin \theta (x)\cdot \hat{y},
\;\; \mbox{as} \;\; |\vec{x}| \rightarrow \infty.
\end{eqnarray}
It is obvious that they have fractional winding number in the $O(3)$-pseudospin
space, and we call them meron.
A pair of $(-1,0)$ and $(0,-1)$ merons form a skyrmion with unit topological 
number.\footnote{More precisely, we should introduce pseudospin operator
with unit length $\hat{m}\equiv \vec{m}/|\vec{m}|$.
Then, topological charge density is given as $q(x)={1\over 8\pi}\epsilon_{ij}
\hat{m}\cdot [\partial_i\hat{m}\times \partial_j\hat{m}]$.}

Let us introduce a couple of quantities;  
soliton charge $Q$ is defined as 
\begin{eqnarray}
Q&\equiv& \int d^2x\Big(\bar{\Psi}\Psi-\bar{\Psi}_0\Psi_0\Big) \nonumber  \\
&=&\int d^2x {1\over 2p}(\epsilon_{ij}\partial_ia^+_j-\epsilon_{ij}\partial_i
a^+_{j,0}) \nonumber   \\
&=&{1\over 2p} \oint dx^j(a^+_j-a^+_{j,0})  \nonumber  \\
&\equiv& {\Phi_{\mbox{\small cs}} \over 2p},
\label{Tcharge}
\end{eqnarray}
where $\Psi_0$ etc., is the quantities of the ground state given by (\ref{groundcon}),
and $\Phi_{\mbox{\small cs}}$ is the total CS flux of soliton.
The electric charge $Q_E$ is $Q_E=-eQ$.
Similarly we define 
\begin{eqnarray}
\bar{\Phi}_{\mbox{\small cs}}&\equiv &\oint dx^j\; a^-_j  \nonumber  \\
\bar{Q}&\equiv &{\bar{\Phi}_{\mbox{\small cs}} \over 2q}.
\label{Tbar} 
\end{eqnarray}
Total angular momentum is given by
\begin{eqnarray}
J&=&\int d^2x \epsilon_{ij}x_i{\cal P}_j,  \label{angular}  \\
{\cal P}_j&=&{1\over 2i}\Big[ \Big\{\bar{\psi}_{\uparrow}D^{\uparrow}_j\psi_{\uparrow}
-\bar{D}^{\uparrow}_j\bar{\psi}_{\uparrow}\cdot\psi_{\uparrow}\Big\} + \Big\{\bar{\psi}_{\downarrow}D^{\downarrow}_j\psi_{\downarrow}
-\bar{D}^{\downarrow}_j\bar{\psi}_{\downarrow}\cdot\psi_{\downarrow}\Big\}\Big].
\end{eqnarray}
It is useful to decompose the angular momentum as $J=U+V$,
\begin{eqnarray}
U&=&{1\over 2i}\epsilon_{ij}\sum_{\sigma} \int d^2x\; x_i(\bar{\psi}_{\sigma}
\partial_j\psi_{\sigma}-\partial_j\bar{\psi}_{\sigma}\psi_{\sigma}),  \nonumber  \\
V&=&-\epsilon_{ij}\int d^2x\; x_i \Big(a^+_j\bar{\Psi}\Psi-a^-_j\bar{\Psi}\sigma_3\Psi
\Big).
\label{UV}
\end{eqnarray}
The CS constraint (\ref{CScon}) is solved in the Coulomb gauge as follows,
\begin{eqnarray}
a^+_j(x)&=&-{p \over \pi}\epsilon_{jk}\int d^2y{(x-y)^k \over |\vec{x}-\vec{y}|^2}
\cdot\bar{\Psi}\Psi(y),  \nonumber   \\
a^-_j(x)&=&-{q \over \pi}\epsilon_{jk}\int d^2y{(x-y)^k \over |\vec{x}-\vec{y}|^2}
\cdot\bar{\Psi}\sigma_3\Psi(y).
\label{CSsol}
\end{eqnarray}
From (\ref{CSsol}),
\begin{equation}
V=-{p\over 2\pi}\Big\{\int d^2x\bar{\Psi}\Psi\Big\}^2-{q\over 2\pi}\Big\{\int d^2x
\bar{\Psi}\sigma_3\Psi\Big\}^2.
\label{V}
\end{equation}
On the other hand for $U$, from (\ref{para1}) and (\ref{fieldeq0})
\begin{equation}
U=\sum_{\sigma}n_{\sigma}\int d^2x\; |\psi_{\sigma}|^2.
\label{U}
\end{equation}
Total angular momentum of soliton, $J(\mbox{soliton})$, is naturally defined as 
\begin{eqnarray}
J(\mbox{soliton})&=&J[\psi_{\sigma}]-
J[\psi_{\sigma,0}=\sqrt{\bar{\rho}/2}]   \nonumber  \\
&=&\sum_{\sigma}n_{\sigma}\int d^2x|\psi_{\sigma}|^2 
-{p\over 2\pi}\Big[\Big\{\int d^2x\; \bar{\Psi}\Psi\Big\}^2-\Big\{\int d^2x\; 
\bar{\Psi}_0\Psi_0\Big\}^2
\Big]  \nonumber   \\
&& \;-{q\over 2\pi}\Big[\Big\{\int d^2x\; \bar{\Psi}\sigma_3\Psi\Big\}^2-\Big\{\int d^2x\; 
\bar{\Psi}_0\sigma_3\Psi_0\Big\}^2\Big].
\label{Jsol}
\end{eqnarray}
Similarly, energy of soliton is given as
\begin{eqnarray}
E&=&\int d^2x\Big\{{1\over 2M}\sum_{\sigma}|(D^{\sigma}_1-iD^{\sigma}_2)
\psi_{\sigma}|^2+{\omega_c\over 2}\bar{\Psi}\Psi+W[\psi_{\sigma}] \Big\} \nonumber  \\
&& \; -\int d^2x\Big\{{\omega_c\over 2}\bar{\Psi}_0\Psi_0+W[\psi_{\sigma,0}] \Big\} \nonumber \\
&=&\int d^2x\Big\{W[\psi_{\sigma}]-W[\psi_{\sigma,0}]\Big\},
\label{E}
\end{eqnarray}
where we have used the self-dual equation (\ref{self}) and the fact that the total number
of electrons is conserved.
For general configuration $\hat{\psi}_{\sigma}(r)e^{in_{\sigma}\theta}$ of the winding numbers 
$(n_1,n_2)$ and the CS gauge fields of the following form
\begin{equation}
a^{\pm}_k=\partial_k\theta\cdot a^{\pm}(r), \;\; 
a^{\uparrow\downarrow}_k=a^+_k\pm a^-_k=\partial_k\theta\cdot a^{\uparrow\downarrow}(r),
\label{apm}
\end{equation}
the energy is explicitly given as 
\begin{eqnarray}
E&=&2\pi \int {dr\over r}\Big[{1\over 2M}\sum_{\sigma}
\Big\{\Big(r{d\hat{\psi}_{\sigma} \over dr}\Big)^2-2r(a^{\sigma}-{eB\over 2}r^2
-n_{\sigma})\hat{\psi}_{\sigma}\Big({d\hat{\psi}_{\sigma}\over dr}\Big)  \nonumber  \\
&&+(a^{\sigma}-{eB\over 2}r^2-n_{\sigma})^2(\hat{\psi}_{\sigma})^2\Big\}
+{\omega_c\over 2}r^2(\bar{\hat{\Psi}}\hat{\Psi}-\bar{\Psi}_0\Psi_0)
+r^2\Big(W[\hat{\psi}_{\sigma}]-W[\hat{\psi}_{\sigma,0}]\Big)\Big]. \nonumber
\end{eqnarray}
Therefore, for energy to be finite, the CS gauge fields must satisfy\footnote{For the 
$(m,m,m)$ state, there is only single CS gauge field $a^+_{\mu}$
and $a^-_{\mu}=0$.
Then, merons have logarithmically divergent energy\cite{moon}.}
\begin{eqnarray}
&& \lim_{r\rightarrow  \infty}a^+(r)={eB\over 2}r^2+{1\over 2}(n_1+n_2),  \nonumber  \\
&& \lim_{r\rightarrow  \infty}a^-(r)={1\over 2}(n_1-n_2).
\label{asymCS}
\end{eqnarray}
From (\ref{asymCS}),
\begin{eqnarray}
&& Q={\pi \over p}\Big({n_1+n_2 \over 2}\Big), 
\;   \bar{Q}=\int d^2x\Big( \bar{\Psi}\sigma_3\Psi -\bar{\Psi}_0\sigma_3\Psi_0\Big)
= {\pi \over q}\Big({n_1-n_2 \over 2}\Big),       \nonumber  \\
&& J(\mbox{soliton})={1\over 2}\Big\{{\pi \over p}\Big({n_1+n_2\over 2}\Big)^2+
{\pi \over q}\Big({n_1-n_2\over 2}\Big)^2\Big\}.
\label{QJ}
\end{eqnarray}
The above $J(\mbox{soliton})$ should be interpreted as spin of solitons, $S$.
On the other hand, statistical parameter of soliton $\alpha(\mbox{soliton})$ is given as follows 
by the argument of the Aharonov-Bohm effect,
\begin{eqnarray}
& & \alpha(\mbox{soliton})=\alpha_++\alpha_-,   \nonumber  \\
& & \alpha_+=-{Q\Phi_{\mbox{\small cs}} \over 2}=-pQ^2,  \nonumber  \\
& & \alpha_-=-{\bar{Q}\bar{\Phi}_{\mbox{\small cs}} \over 2}=-q\bar{Q}^2.
\label{alphasol}
\end{eqnarray}
Therefore, the spin-statistics relation, $S=-{1\over 2\pi}\alpha$, 
holds generally for solitons in the present system.
Actually, it is obvious from the above consideration that there are two CS gauge fields,
$a^+_{\mu}$ and  $a^-_{\mu}$, in the present model and they contribute to
both spin and statistics {\em additively}.
This result is rederived in later discussion in Sec.4 which is based on an effective theory
of solitons.

Let us turn to the numerical solutions to the field equations (\ref{fieldeq2}).
For example for the $(-1,-1)$ vortex, asymptotic solution near the center 
of soliton is obtained as 
\begin{equation}
v_i(\l) \sim \beta_i-{\l^2 \over 2},
\label{asym}
\end{equation}
where $\beta_i$'s are parameters to be determined by the boundary condition
at spatial infinity.
Similar asymptotic solutions are obtained for general $(n_1,n_2)$.
Using these asymptotic solutions, we obtain numerical solutions.

In the DL FQHE, there are more than one QHS's at the same filling factor.
Actually as we showed in Sec.2, the present CS gauge theory describes
QHS at $\nu=\pi/p$, which is independent of the parameter $q$.
The value of $q$ is determined by the inter and intra-layer Coulomb interactions
$W[\psi_{\sigma}]$, which depend on inter-layer distance $d$.
For each $d$, some specific Halperin state is a good variational wave function.

As an example, we shall consider the case $\nu={2\over 5}$,
and at this filling fraction, Halperin's states $(3,3,2)$, $(5,5,0)$ and $(1,1,4)$
are possible candidates for the QHS.
From (\ref{halperin}), it is obvious that the state $(5,5,0)$ corresponds to such a case
that the inter-layer distance $d$ is sufficiently large and the inter-layer 
Coulomb interaction is negligibly small.
On the other hand for the $(1,1,4)$ state, the inter-layer Coulomb interaction
dominates over the intra-layer interaction.
Though this situation does not seem to be realized in samples in experiment so far,
some interesting features of solitons appear in this QHS as we shall see.
The state $(3,3,2)$ is in between.

Vortex solution $(n_1=-1, n_2=-1)$ in the $(3,3,2)$ state is given in Fig.2.
The picture shows that size of vortex is about four times of the magnetic length $l_0
=1/\sqrt{eB}$, and its form is quite similar to that in the single-layer system\cite{vortex}.
Meron solutions $(n_1=-1,n_2=0)$ etc. are also given in Fig.2.
It should be remarked that the density in the lower layer {\em increases} at the center
of the meron.
It is a result of the inter-layer Coulomb interaction.
Size of the meron is about $4.5\sqrt{2}l_0$, and it is larger than the vortex.
Creation energies of vortex and merons are calculation from these solutions
and (\ref{E}).
We also show the numerical solutions of $(-2,0)$ and $(-2,-1)$ solitons
in Fig.2.

Similarly, meron solution for $(5,5,0)$ state is shown in Fig.3.
As there is no inter-layer interaction, the density in the lower layer
is not influenced by the deviation of density in the lower layer.
Finally for the $(1,1,4)$ state, meron solution in Fig.4 has a long undulating tail
because of the dominant inter-layer interaction.

From the above meron solutions, we can calculate normalized pseudospin vector
$\hat{m}(x)=\vec{m}(x)/|\vec{m}(x)|$.
Merons have a fractional topological charge in the $O(3)$ space.
In the following section, we shall derive an effective field theory of solitons,
and show which field theory describes $\hat{m}(x)$ and how the topological
charge is related with electric charge of solitons.
As we see, the lowest Landau level projection plays a very important role there.

Finally, we shall give wave functions corresponding to the topological
solitons in the first-quantization language.
For soliton of $(n_1,n_2)$ type which is located at $(w, \bar{w})$,
\begin{eqnarray}
\Psi_{S}(z^{\uparrow};z^{\downarrow};w)&=&\prod_{i<j}(z^{\uparrow}_i-z^{\uparrow}_j)^m
\prod_{i<j}(z^{\downarrow}_i-z^{\downarrow}_j)^m
\prod_{i,j}(z^{\uparrow}_i-z^{\downarrow}_j)^n  \nonumber  \\
&& \; \times \prod_i(z^{\uparrow}_i-w)^{-n_1}\prod_i(z^{\downarrow}_i-w)^{-n_2} \nonumber  \\
&&\; \times \exp \Big(-{eB\over 4}(\sum |z^{\uparrow}_i|^2+\sum |z^{\downarrow}_i|^2)\Big).
\label{halperin2}
\end{eqnarray}

%%%%%%%%%%%%%%%%%%%%%%%%%%%%%%%%%%%%%%%%%%%%%%%

\section{Effective action of vortices and merons}
\setcounter{equation}{0}

In this section, we shall derive an effective field theory of topological solitons
considered in Sec.3.
At specific filling factor (\ref{ground}), these solitons are excitations of low energies.
Lowest Landau level condition is expressed in terms of topological current
of these solitons, as we shall see in later discussion.

We first prameterize the bosonized electron fields as follows\cite{lee},
\begin{eqnarray}
\psi_{\sigma}&=&J^{1/2}_0\phi \; z_{\sigma},  \nonumber  \\
 \phi&=&e^{i\tilde{\theta}}\phi_{v}\;\in\;  U(1), \;\; z=(z_1 \;\; z_2)^t\; \in \; CP^1,
\label{para}
\end{eqnarray}
where $e^{i\tilde{\theta}}$ is a regular part and $\phi_v$ is a topologically nontrivial part, 
which represents vortex degrees of freedom.
$CP^1$ variable $z_{\sigma}$ represents pseudospin degrees of freedom and satisfies
$CP^1$ condition $\sum |z_{\sigma}|^2=1$, and $\psi_{\sigma}$ is
invariant under $z_{\sigma}\rightarrow e^{i\alpha}z_{\sigma}$ and 
$\phi \rightarrow e^{-i\alpha}\phi$.

Substituting (\ref{para}) into (\ref{Lpsi}),
\begin{eqnarray}
{\cal L}_{\psi}&=&-J_0\Big[\bar{\phi}\partial_0\phi+(\bar{z}\partial_0 z)-ia^+_0-ia^-_0
(\bar{z}\sigma_3z)\Big]  \nonumber   \\
&&-{J_0 \over 2M}\Big[\bar{\phi}\partial_j\phi+\bar{z}\partial_j z-i(a^+_j-eA_j)-ia^-_j
(\bar{z}\sigma_3z)\Big]   \nonumber  \\
&& \; \times \Big[\partial_j\bar{\phi}\cdot\phi+\partial_j\bar{z}\cdot z+i(a^+_j-eA_j)
+ia^-_j(\bar{z}\sigma_3z)\Big] +{\cal L}_z,  \nonumber  \\
{\cal L}_z&=&-{J_0\over 2M}\Big[\overline{D_jz}\cdot D_jz+(\bar{z}D_jz)^2\Big], 
\label{Lpsi2}
\end{eqnarray}
where 
\begin{equation}
D_{\mu}z=(\partial_{\mu}-ia^-_{\mu}\sigma_3)z,
\label{covd}
\end{equation}
and we have neglected derivative terms of $J_0$, for $J_0$ is {\em not} a dynamical
variable\footnote{It is obvious that the term $J^{1/2}_0\partial_0 J^{1/2}_0$ is total derivative} 
and its value is determined by LLL condition as we shall see.
We have also neglected the interaction term $V[\psi_{\sigma}]$, which 
will be discussed later on.

We shall perform duality transformation\cite{zhang}. 
First, we use the following identity;
\begin{equation}
e^{-\int d\tau d^2x {K\over 2} \varphi_j\varphi_j} =\int [DJ]
e^{-\int d\tau d^2x ({1\over 2K}J_jJ_j+iJ_j\varphi_j)}.
\end{equation} 
In the present case, $K={\bar{\rho} \over M}$ and 
\begin{equation}
\varphi_j=\bar{\phi}{\partial_j \over i}\phi+\bar{z}\cdot {\partial_j \over i}z
-(a^+_j-eA_j)-a^-_j(\bar{z}\sigma_3z),
\end{equation}
and the Lagrangian ${\cal L}$ is expressed as 
\begin{eqnarray}
{\cal L}&=&-J_0\Big[\bar{\phi}\partial_0 \phi+(\bar{z}\partial_0 z)-ia^+_0
-ia^-_0(\bar{z}\sigma_3 z)\Big]-{1 \over 2K} \vec{J}\cdot \vec{J}-iJ_j
\Big[\bar{\phi}{\partial_j \over i}\phi+\bar{z}\cdot {\partial_j \over i}z\Big]  \nonumber  \\
&&+iJ_j(a^+_j-eA_j)+iJ_ja^-_j(\bar{z}\sigma_3z)+{\cal L}_z+{\cal L}_{\mbox{\small CS}}.
\label{L2}
\end{eqnarray}

It is obvious that  integration over $\tilde{\theta}$ gives the divergentless condition
\begin{equation}
\partial_{\mu}J_{\mu}=0.
\label{divless}
\end{equation}
Solution to (\ref{divless}) is 
\begin{equation}
J_{\mu}={1\over 2\pi}\epsilon_{\mu\nu\lambda}\partial_{\nu}b_{\lambda}.
\label{bfield}
\end{equation}
Substituting (\ref{bfield}) into (\ref{L2}),
\begin{eqnarray}
{\cal L}&=&-{1\over 2\pi}\epsilon_{ij}\partial_ib_j
\Big[\bar{\phi}_v\partial_0 \phi_v+(\bar{z}\partial_0 z)-ia^+_0
-ia^-_0(\bar{z}\sigma_3 z)\Big]  \nonumber   \\
&&-{1\over 2K}{1\over (2\pi)^2}(\partial_0b_j-\partial_jb_0)^2
-{i\over 2\pi}\epsilon_{ij}(\partial_0b_i-\partial_ib_0)
\Big[\bar{\phi}_v{\partial_j \over i}\phi_v+\bar{z}\cdot {\partial_j \over i}z\Big]  \nonumber   \\
&&+{i\over 2\pi}\epsilon_{ij}(\partial_0b_i-\partial_ib_0)(a^+_j-eA_j+a^-_j(\bar{z}\sigma_3z))
+{\cal L}_z+{\cal L}_{\mbox{\small CS}}.
\label{L3}
\end{eqnarray}

Next we integrate over $b_0$ with the Coulomb gauge $\partial_jb_j=0$.
The form of the integrand is the following,
\begin{eqnarray}
&&-{1\over 2K}{1\over (2\pi)^2}(\partial_j b_0)^2+ib_0\Phi,  \nonumber   \\
\Phi&=&-{1 \over 2\pi}\epsilon_{ij}\partial_i
\Big[\bar{\phi}_v{\partial_j \over i}\phi_v+\bar{z}\cdot {\partial_j \over i}z
-(a^+_j-eA_j)-a^-_j(\bar{z}\sigma_3z)\Big].
\end{eqnarray}
Integration gives 
\begin{equation}
\int [Db_0]e^{\int d^3x[-C(\partial_jb_0)^2+ib_0\Phi]}
=e^{-{1\over 4}\int\Phi {1\over C}\Delta^{-1}\Phi},
\end{equation}
where $C={1\over 2K(2\pi)^2}$ and 
\begin{equation}
\int\Phi \Delta^{-1}\Phi=-{1\over 2\pi} \int d\tau\int d^2xd^2y \; \Phi(x)
\ln |\vec{x}-\vec{y}| \Phi(y),
\end{equation}
\begin{equation}
\Delta^{-1}(x)=\int {d^2k \over (2\pi)^2}{e^{ik\cdot x} \over k^2}.
\end{equation}
We furthermore integrate over the CS fields $a^{+}_0$ which gives the CS constraint,
\begin{equation}
\epsilon_{ij}\partial_ia^+_j=2p|\Psi|^2=2pJ_0={p\over \pi}\epsilon_{ij}\partial_ib_j,
\label{CSa+}
\end{equation}
and $\Phi$ is expressed as
\begin{equation}
\Phi=-{1\over 2\pi}\epsilon_{ij}\partial_i\Big[\bar{\phi}_v{\partial_j \over i}\phi_v+
\bar{z}\cdot {D_j \over i}z-({p\over \pi}b_j-eA_j)\Big].
\end{equation}

Let us consider quantization of the field $b_j$.
Kinetic term is given as $(\partial_0 b_j)^2$ and a system is simply a harmonic oscillator
with the following form of the action $L=\dot{x}^2+(x-a)^2$.
We shall consider static configuration of solitons.
Then, Largangian of $b_j$ field, ${\cal L}_b$, is given by
\begin{equation}
 {\cal L}_b=-C (\partial_0 b_j)^2-{1\over 4C}\int d^2y \Phi (x,\tau)\Delta^{-1}(x-y)\Phi (y,\tau).
\label{Lb}
\end{equation}
It is convenient to take $B(x)\equiv \epsilon_{ij}\partial_ib_j$ as dynamical variable
for quantization in stead of $b_j(x)$ itself which is subject to the Coulomb gauge
condition.
The original field $b_j$ is expressed in terms of $B$ as 
\begin{equation}
b_i(x)=\epsilon_{ij}\partial^x_j\int d^2y \Delta^{-1}(x-y)B(y).
\label{bB}
\end{equation}
We shall introduce Fourier transformed variables as 
\begin{eqnarray}
b_i(x)&=&\int {d^2q \over (2\pi)^2} e^{iq\cdot x}\tilde{b}_i(q),  \nonumber   \\
B(x)&=&\int {d^2q \over (2\pi)^2} e^{iq\cdot x}\tilde{B}(q), \; \mbox{etc.}
\label{Fou}  
\end{eqnarray}
In terms of Fourier transformed fields,
\begin{eqnarray}
&&\int d^2 x \; (\partial_0 b_j)^2=\int {d^2q \over (2\pi)^2}{1 \over q^2}\partial_0
\tilde{B}(q)\partial_0\tilde{B}(-q),  \nonumber  \\
&&\int d^2xd^2y \Phi(x)\Delta^{-1}(x-y)\Phi(y)=\int {d^2q \over (2\pi)^2}
\tilde{\Phi}(q){1\over q^2}\tilde{\Phi}(-q).
\label{Lagfourier}
\end{eqnarray}
Then, returning to the real-time formalism as $t=-i\tau$,
canonical conjugate variable of $\tilde{B}(q)$ is
\begin{equation}
\Pi_{B}(q)=C\;{2 \over (2\pi)^2 q^2}\partial_t\tilde{B}(-q).
\label{piB}
\end{equation}
Hamiltonian is given as 
\begin{equation}
H_B=\int d^2q \Big[{K \over 2}q^2(2\pi)^4\Pi_B(q)\Pi_B(-q)+{K\over 2}\tilde{\Phi}(q){1\over q^2}
\tilde{\Phi}(-q)\Big],
\label{H}
\end{equation}
and 
\begin{equation}
\tilde{\Phi}(q)={1\over 2\pi}{1\over \nu}\tilde{B}(q)+\cdot \cdot\cdot.
\label{tilPhi}
\end{equation}
From (\ref{H}) and (\ref{tilPhi}), it is obvious that the ground state is the 
configuration which satisfies $\Phi=0$ and excitation energies are given
by $N\omega_c$,
where $N$ is the number of excitations and $\omega_c=eB/M$. 
 
Ground state condition $\Phi=0$ is nothing but the LLL condition, which
is given as follows\cite{sondhi},
\begin{equation}
J^v_0+J^S_0-{2p \over (2\pi)^2}\epsilon_{ij}\partial_ib_j+{e\over 2\pi}
\epsilon_{ij}\partial_iA_j=0,
\label{LLL}
\end{equation}
where topological densities and currents are defined as
\begin{eqnarray}
J^v_{\mu}&=&{1\over 2\pi}\epsilon_{\mu\nu\lambda}\partial_{\nu}\Big(\bar{\phi}_v
{\partial_{\lambda} \over i}\phi_v\Big),   \label{Jv}  \\
J^S_{\mu}&=&{1\over 2\pi}\epsilon_{\mu\nu\lambda}\partial_{\nu}\Big(\bar{z}
{D_{\lambda} \over i}z\Big).   \label{JS} 
\end{eqnarray}
There is another $U(1)\otimes U(1)$ gauge-invariant topological current, 
which is defined as 
\begin{equation}
J^3_{\mu}={1\over 2\pi}\epsilon_{\mu\nu\lambda}\partial_{\nu}\Big(\bar{z}\sigma_3
{\tilde{D}_{\lambda} \over i}z\Big),
\label{J3}
\end{equation}
where $\tilde{D}_{\l}=\partial_{\l}-i(a^+_{\l}-eA_{\l})$.
This topological charge distinguishes between $(-1,0)$ and $(0,-1)$ merons. 

In the small-size or the long-distance limit of the topological solitons, 
the above currents are given as 
\begin{eqnarray}
J^v_0(x)&=&\sum_{\alpha}q^v_{\alpha}\delta (x-x^{v,\alpha}), \nonumber   \\
J^v_j(x)&=&\sum_{\alpha}q^v_{\alpha}\dot{x}^{v,\alpha}_j\delta (x-x^{v,\alpha}), \nonumber   \\
J^S_0(x)&=&\sum_{\alpha}q^S_{\alpha}\delta (x-x^{S,\alpha}), \nonumber   \\
J^S_j(x)&=&\sum_{\alpha}q^S_{\alpha}\dot{x}^{S,\alpha}_j\delta (x-x^{S,\alpha}), 
\label{topcurrent}
\end{eqnarray}
where index $\alpha$ distinguishes solitons and
$q^{v}_{\alpha}$ and $q^{S}_{\alpha}$ are topological charges of solitons, i.e.,
\begin{eqnarray}
&&q^v=-1 \;\; \mbox{for} \;\; (-1,-1)\; \mbox{vortex}, \nonumber    \\
&&q^S=-{1\over 2} \;\; \mbox{for} \;\; (-1,0), (0,-1)\; \mbox{merons},  \nonumber   \\
&&q^S=+{1\over 2} \;\; \mbox{for} \;\; (1,0), (0,1)\; \mbox{merons},  
\label{topnumber}
\end{eqnarray}
and $x^{v,\alpha}$ and $x^{S,\alpha}$ are their coordinates.
From (\ref{para2}), merons $(1,0)$ etc. do not exist as a regular solution
to the self-dual equation.
This situation is similar to the single-layer system, in which quasi-particle excitation
does not exist as a regular solution.
However, these solitons must exist as low-energy excitations and their
behavior away from the center is described by (\ref{para2}).

The LLL condition (\ref{LLL}) determines the density of electrons in terms of
the topological currents.
For example from (\ref{CSa+}), the vortex $(-1,-1)$ has charge $\nu e$
and the merons $(-1,0)$ and $(0,-1)$ have charge ${\nu e \over 2}$.

Here, a few remarks about the LLL condition are in order.
The condition (\ref{LLL}) is correct for static configurations.
For moving soliton, it is not so difficult to show that
there appears additional term which is proportional to
$\epsilon_{ij}\partial_i(J^v_j+J^S_j)$.
Existence of this term, however, does not change the discussion on 
statistics of solitons, etc, given below.
For example, 
\begin{equation}
\int_{S_0} d^2x\;\epsilon_{ij}\partial_iJ_j=\int_{\partial S_0}J_jdx_j,
\end{equation}
where $S_0$ is a domain and $\partial S_0$ is its boundary.
For sufficiently large domain whose boundary is far away from solitons, 
the above integral is vanishingly small and therefore charge of solitons
is determined solely by the topological density.

There is obviously an ambiguity in the decomposition (\ref{para}), i.e.,
the bosonized electron fields are invariant under the following transformation,
\begin{equation}
\phi \rightarrow e^{-i\varphi}\phi,  \; z_{\sigma} \rightarrow e^{i\varphi}z_{\sigma},
\label{singular}
\end{equation}
where the phase $\varphi$ can have a {\em nontrivial} winding number.
This singular transformation is allowed at the center of solitons with topological numbers
$n_1\neq 0$ and $n_2\neq 0$, as the amplitude $J_0$ vanishes there.
From the definition of the topological charges (\ref{Jv}) and (\ref{JS}),
\begin{eqnarray}
J^v_0 &\rightarrow& J^v_0-{1\over 2\pi}\epsilon_{ij}\partial_i\partial_j\varphi, \nonumber  \\
J^S_0 &\rightarrow& J^S_0+{1\over 2\pi}\epsilon_{ij}\partial_i\partial_j\varphi.
\label{JvJS}
\end{eqnarray}
Therefore, the total topological charge is invariant under (\ref{singular}).

With the LLL condition (\ref{LLL}), the Lagrangian is given as 
\begin{eqnarray}
{\cal L}&=&-{1\over 2\pi}\epsilon_{ij}\partial_ib_j\Big[\bar{\phi}_v\partial_0\phi_v
+(\bar{z}\partial_0z)-ia^-_0\cdot (\bar{z}\sigma_3z)\Big]-{i\over 2\pi}\epsilon_{ij}\partial_0b_i
\Big[\bar{\phi}_v{\partial_j \over i}\phi_v+\bar{z}\cdot {\partial_j \over i}z\Big] \nonumber  \\
&&+\; {i\over 2\pi}\epsilon_{ij}\partial_0b_i\cdot (a^+_j-eA_j)+{i\over 2\pi}\epsilon_{ij}
\partial_0b_i\; a^-_j\cdot (\bar{z}\sigma_3z)+{\cal L}_z+{\cal L}_{\mbox{\small CS}}(a^-_{\mu}).
\label{Leff}
\end{eqnarray}
After some calculation, it is shown 
\begin{equation}
\mbox{1st+2nd+4th terms of (\ref{Leff})}=-ib_i  \cdot (J^v_i+J^S_i).
\label{Hopf}
\end{equation}
On the other hand, the third term is evaluated as follows;
\begin {equation}
{i \over 2\pi}\epsilon_{ij} \partial_0b_i\cdot (a^+_j-eA_j)  
\; \Rightarrow -{i\over 2\pi}\epsilon_{ij}b_i\cdot\partial_0 (a^+_j-eA_j).
\label{third}
\end{equation}
Let us denote $\hat{J}_{\mu}=J^v_{\mu}+J^S_{\mu}$.
In the Coulomb gauge,
\begin{equation}
(a^+_j-eA_j)(x)=2\pi \int d^2y \epsilon_{jk} \partial^x_k \Delta^{-1}(x-y) \; \hat{J}_0(y).
\label{sol1}
\end{equation}
By the conservation of the topological currents 
\begin{equation}
\partial_{\mu}\hat{J}_{\mu}=0,
\end{equation}
\begin{eqnarray}
\epsilon_{ij}\partial_0(a^+_j-eA_j)(x)&=&-2\pi \int d^2y \partial^x_i
\Delta^{-1}(x-y)\;\partial_0\hat{J}_0(y)  \nonumber   \\
&=&2\pi\int d^2y \partial^x_i\partial^x_k\Delta^{-1}(x-y)\cdot \hat{J}_k(y).
\label{sol2}
\end{eqnarray}
On the other hand, 
\begin{equation}
b_i(x)={2\pi^2 \over p} \int d^2y \epsilon_{ij}\partial^x_j\Delta^{-1}(x-y)\cdot \hat{J}_0(y)
+{\pi \over p}eA_i(x).
\label{sol3}
\end{equation}
Then from (\ref{sol2}) and (\ref{sol3})
\begin{eqnarray}
\int d^2x b_i(x)\epsilon_{ij}\partial_0(a^+_j-eA_j)(x) &=& 
{2\pi^2 \over p}\int d^2xd^2yd^2z \Big[\partial^x_i\partial^x_k\Delta^{-1}(x-z)\cdot \hat{J}_k(z)
\epsilon_{ij}\partial^x_j
\Delta^{-1}(x-y)  \nonumber  \\
&& \; \times \hat{J}_0(y)\Big]   
+ {\pi \over p}\int d^2xd^2zeA_i(x)\partial^x_i\partial^x_k\Delta^{-1}(x-z)
\cdot \hat{J}_k(z)  \nonumber  \\
&=&0,
\end{eqnarray}
where we have used the Coulomb gauge condition.

Therefore, the effective Lagrangian ${\cal L}_E$ has the following final form;
\begin{equation}
{\cal L}_E={\cal L}_z+{\cal L}_{\mbox{\small CS}}(a^-_{\mu})-ib_i\cdot \hat{J}_i.
\label{effeL}
\end{equation}
With the LLL condition (\ref{LLL}), 
the term (\ref{Hopf}) is nothing but the Hopf term which determines 
statistics of solitons\cite{hopf}.
For vortex with $(-1,-1)$, $J^v_{\mu}$ has charge $-1$. 
It is not difficult to show that statistical parameter of that excitation coming from
the above Hopf term is $\pi\nu$, where the filling factor
$\nu$ is given by $\nu={2\pi\bar{\rho} \over eB}={\pi \over p}$.  
On the other hand for merons $(-1,0)$ and $(0,-1)$, $J^S_{\mu}$ has charge $-1/2$,
and then the statistical parameter from the Hopf term is $\pi\nu ({1\over 2})^2$. 
The above results are consistent with the discussion in terms of the 
Aharonov-Bohm effect in Sec.3, i.e., the Hopf term in (\ref{effeL}) gives the
statistical parameter $\alpha_+$ in (\ref{alphasol}).

The above Hopf term also gives spin to solitons\cite{hopf}, which is consistent with
the spin-statistics relation.
As we showed in Sec.3, merons in the present system have the extra terms
both in their spin and statistical parameter, besides those coming from the Hopf term.
These terms come from the coupling with the CS field $a^-_{\mu}$, as we can see
from (\ref{effeL}) and the discussion in Sec.3.
Actually recently,  CS gauge theory, which is a relativistic version of (\ref{effeL})
{\em without} Hopf term,
is studied\cite{rela}.
There, it is shown that solitons similar to those in the present system have a
fractional spin and also fractional statistics.
We also studied topological solitons in the effective theory (\ref{effeL})
very recently\cite{IS2} and found that the CS gauge field $a^-_{\mu}$
gives spin and statistical parameter to merons which are consistent with the results in
Sec.3.

In the following section, we shall consider the quantization of  multi-soliton configurations
retaining only center coordinates of solitons.
In that discussion, it will become clear that the above Hopf term surely
determines statistics of solitons.

%%%%%%%%%%%%%%%%%%%%%%%%%%%%%%%%%%%%%%%%%%%%%%%%%

\section{Quantum mechanics of topological excitations}
\setcounter{equation}{0}

In the previous section, we have obtained the effective Lagrangian of topological
solitons (\ref{effeL}) by the duality transformation.
Also as we showed by the numerical calculation in Sec.3, topological solitons 
have a finite size of order of the magnetic length.
Actually form of the meron is determined by the terms ${\cal L}_z+
{\cal L}_{\mbox{\small CS}}(a^-_{\mu})$
in ${\cal L}_E$, and long-wave-length properties of topological excitations are determined
by the term $b_i\cdot \hat{J}_i$.
Especially, by using the topological currents in the form of the small-size limit 
(\ref{topcurrent}), one can easily obtain an effective Largangian of multi-soliton
system from $b_i\cdot \hat{J}_i$, which retaines only coordinates
of solitons as collective coordinates.

The LLL condition is solved as (\ref{sol3}). 
It is not so difficult to calculate the long-distance limit of the term $b_i\cdot \hat{J}_i$,
\begin{eqnarray}
\int d^2x b_i(x)\cdot \hat{J}_i(x)&=&
\pi\bar{\rho}\epsilon_{ij}\Big(\sum_{\alpha}q^v_{\alpha}\dot{x}^{v,\alpha}_ix^{v,\alpha}_j
+\sum_{\alpha}q^S_{\alpha}\dot{x}^{S,\alpha}_ix^{S,\alpha}_j\Big)  \nonumber  \\
&&-\nu\Bigg(\sum_{\alpha\neq\beta}q^v_{\alpha}q^v_{\beta}\epsilon_{ij}\dot{x}^{v,\alpha}_i
{(x^{v,\alpha}-x^{v,\beta})_j \over (x^{v,\alpha}-x^{v,\beta})^2}  
+\sum_{\alpha,\beta}q^S_{\alpha}q^v_{\beta}\epsilon_{ij}\dot{x}^{S,\alpha}_i
{(x^{S,\alpha}-x^{v,\beta})_j \over (x^{S,\alpha}-x^{v,\beta})^2}  \nonumber   \\
&&+\sum_{\alpha,\beta}q^v_{\alpha}q^S_{\beta}\epsilon_{ij}\dot{x}^{v,\alpha}_i
{(x^{v,\alpha}-x^{S,\beta})_j \over (x^{v,\alpha}-x^{S,\beta})^2}
+\sum_{\alpha\neq\beta}q^S_{\alpha}q^S_{\beta}\epsilon_{ij}\dot{x}^{S,\alpha}_i
{(x^{S,\alpha}-x^{S,\beta})_j \over (x^{S,\alpha}-x^{S,\beta})^2}\Bigg).  \nonumber  \\
\label{smallL}
\end{eqnarray}
It is obvious that the terms proportional to $\nu$ in (\ref{smallL}) determine
the statistics of  solitons,
because they are rewritten as\footnote{It should be remarked that in the present
system, the size of solitons or the magnetic length itself plays a role of  short-distance
cutoff.}
\begin{equation}
\sum_{\alpha,\beta}{d \over dt}\theta (x^{\alpha}-x^{\beta}),
\label{theta}
\end{equation}
where $\theta(x)=\arctan \Big({x_2 \over x_1}\Big)$.
In the path-integral formalism, this statistical term gives nontrivial contribution
when world lines of two particles entwine each other.
As we briefly mentioned in Sec.2 and Sec.3, similar terms appear
from the coupling of solitons with the CS gauge field $a^-_{\mu}$.

Anyonic properties of topological excitations plays an important role,
for example, for hierarchical structure of the FQHE.
Anyonic vortices and merons can be expressed in terms of 
boson fields with CS flux attached.
At specific filling factors, these bosonized anyons can Bose condense,
as the bosonized electrons.
These states correspond to daughter states of the FQHE, and it is interesting
to compare theoretical prediction of filling fractions of the FQHS with experimental
observations\cite{IS3}.

In this section, we shall study quantum system of quasi-excitations
in the FHQS.
The first term of (\ref{smallL}) determines canonical commutation 
relations (CCR) of soliton coordinates.
Generally, the quantum system of anyons, whose Lagrangian is given by  
(\ref{smallL}), is very difficult to solve.
Therefore, we shall mostly concentrate on two-particle systems in the rest
of discussion.
In specific but physically interesting cases of two-particle systems,
it can be shown that the statistical terms can be ignored
in the first approximation, as we discuss later on.
On the other hand, we shall take into account interactions between solitons.

Returnning to the real-time formalism, the Lagrangian of the multi-soliton system
is given as,
\begin{eqnarray}
{\cal L}_{\mbox{\small soliton}}&=&-\pi\bar{\rho}\epsilon_{ij}\Big(\sum_{\alpha}q^v_{\alpha}
\dot{x}^{v,\alpha}_ix^{v,\alpha}_j
+\sum_{\alpha}q^S_{\alpha}\dot{x}^{S,\alpha}_ix^{S,\alpha}_j\Big)  \nonumber  \\
&&-\sum_{\alpha}\Big(D(q^v_{\alpha})+D(q^S_{\alpha})\Big)
-\sum_{\omega,\omega'=v,S}\sum_{\alpha,\beta} (e\nu)^2 q^{\omega}_{\alpha}q^{\omega'}_{\beta}
W(x^{\omega,\alpha}-x^{\omega',\beta}),
\label{Lagsol}
\end{eqnarray}
where $D(q^v_{\alpha})=D_v$ and $D(q^S_{\alpha})=D_m$ are energies to create the 
vortex and the meron, respectively (see (\ref{E})), and 
$W(x^{\omega,\alpha}-x^{\omega',\beta})$ is the Coulomb
interactions between solitons.
From (\ref{Lagsol}), the CCR 's among soliton coordinates are
given as\cite{fetter},
\begin{eqnarray}
&& [x^{v,\alpha}_i, x^{v,\beta}_j] = -i {\epsilon_{ij} \over 2\pi\bar{\rho}q^v_{\alpha}}
\delta_{\alpha\beta}, \nonumber   \\
&& [ x^{v,\alpha}_i, x^{S,\beta}_j ] = 0, \nonumber   \\
&& [ x^{S,\alpha}_i, x^{S,\beta}_j ] = -i {\epsilon_{ij} \over 2\pi\bar{\rho}q^S_{\alpha}} 
\delta_{\alpha\beta}.
\label{CCR}
\end{eqnarray}
Here we have neglected the statistical terms.
After obtaining quantum mechanical states of two-particle systems,
we shall give plausible argument supporting this approximation.
From (\ref{CCR}), conjugate momentum of $x_i$ is $\epsilon_{ij}x_j$.
This obviously corresponds to the cyclotron motion in the classical picture.
From this fact, dispersion relation of a pair of solitons is, in some cases, obtained
quite easily. 

Actually concerning vortices, the Lagrangian (\ref{Lagsol}) has a similar structure with 
that of vortices in the single-layer system.
It is rather straightforward to show that a vortex-antivortex pair, which are separated 
with each other more than the magetic length, form a magneto-roton excitation.
Its dispersion relation is given as\cite{zhang} 
\begin{equation}
E(p_t)=2D_v-{\nu^3 e^2 \over p_tl_0^2},
\label{dis1}
\end{equation}
where $p_t$ is the eigenvalue of the total momentum.
This result is obtained by ignoring the statistical terms, but is in qualitative agreement
with the results of the exact diagonalization\cite{haldane}.

Similar relation is obtained for a pair of merons.
However, there is a very important effect which we have to take into account
in the DL electron systems, i.e., inter-layer tunneling.
Hamiltonian of the inter-layer tunneling is given as,
\begin{eqnarray}
{\cal H}_{T}&=&-\Delta_{\mbox{\small SAS}}\Big(\bar{\psi}_{\uparrow}\psi_{\downarrow}
+\bar{\psi}_{\downarrow}\psi_{\uparrow}\Big)  \nonumber   \\
&=&-\Delta_{\mbox{\small SAS}}\Big(\bar{\Psi}\sigma_1\Psi\Big),
\label{sas}
\end{eqnarray}
where $\Delta_{\mbox{\small SAS}}$ is the energy splitting between symmetric
and anti-symmetric states with respect to the pseudospin index.\footnote{Strictly
speaking, the tunneling term (\ref{sas}) should be written in terms of the original
fermionic electron operators.
As a result, an extra factor like $\exp (2i\sigma_2\int^xa^-_jdx_j)$ appears.
This factor is important to preserve the gauge symmetry. 
However, results of the later discussion are not affected by this factor.}

In terms of the parametrization (\ref{para}),
\begin{equation}
{\cal H}_{T}=-\Delta_{\mbox{\small SAS}}J_0(\bar{z}\sigma_1z),
\label{sas2}
\end{equation}
and therefore it is obvious that ${\cal H}_{T}$ gives an linear attractive confining
force between $(-1,0)-(0,-1)$ pair,  $(-1,0)-(1,0)$ pair,
etc., see Fig.5.

Let us first consider a pair of $(-1,0)$ and $(1,0)$ merons, whose coordinates are
$(x_1,y_1)$ and $(x_2,y_2)$, respectively.
From (\ref{CCR}), the nonvanishing CCR's are
\begin{equation}
[x_1,y_1]=iR, \;\; [x_2,y_2]=-iR, \;\; R\equiv {2l_0^2 \over \nu}.
\label{CCR2}
\end{equation}
Introducing the center of mass and relative coordinates as $X={1\over 2}(x_1+x_2), \;
Y={1\over 2}(y_1+y_2), \; x=x_1-x_2$ and $y=y_1-y_2$, the CCR' are
\begin{equation}
[X,y]=iR, \; [Y,x]=-iR.
\label{CCR3}
\end{equation}
From (\ref{CCR3}), it is obvious that $Y=R\hat{p}_x$ and $y=R\hat{p}_X$, 
where $\hat{p}_x$ and $\hat{p}_X$
are relative and total momentum operators, respectively.
On the other hand, potential energy of this system is give as 
\begin{equation}
V_1(r_{12})=-\Big({e\nu\over 2}\Big)^2{1\over r_{12}}+T_1r_{12},
\label{pot1}
\end{equation}
where $r_{12}=|\vec{r}_1-\vec{r}_2|$ and tension $T_1 \propto \Delta_{\mbox{\small SAS}}$.
If the total momentum of the pair is directed to the $x$-axis with magnitude
$\hat{p}_X=p_X$, the energy is given as
\begin{equation}
E(p_X)=2D_m-\Big({e\nu\over 2}\Big)^2{\nu\over 2l_0^2}{1\over p_X}+
{2T_1l_0^2\over \nu}p_X.
\label{E1}
\end{equation}
The above dispersion is correct for $r_{12}>l_0$ and $p_X>l_0/R$.
Therefore, $(-1,0)-(1,0)$ meron pair also has magneto-roton dispersion.
However, in contrast with the vortex-antivortex pair, its energy increase linearly
with total momentum.

Let us turn to the $(-1,0)-(0,-1)$ meron pair.
Nontrivial CCR's are
\begin{equation}
[x_1,y_1]=iR, \;\; [x_2,y_2]=iR.
\label{CCR4}
\end{equation}
From (\ref{CCR4}),
\begin{equation}
[X,Y]={i \over 2}R, \;\; [x,y]=2iR.
\label{CCR5}
\end{equation}
In terms of the complex coordinates $\xi=x+iy$, $[\bar{\xi},\xi]=-4R$, or by introducing
radial and angle coordinates as $\xi=re^{i\varphi}$, the CCR is $[r^2,\varphi ]=-4iR$.
Energy of this system is given as 
\begin{equation}
E(r)=2D_m+\Big({e\nu \over 2}\Big)^2{1\over r}+T_2r,
\label{E2}
\end{equation}
where again $T_2 \propto \Delta_{\mbox{\small SAS}}$.
It is obvious that merons are rotating around the center of mass with a fixed
radius.
The minimum of the energy is located at $r_m={e\nu \over 2}{1\over \sqrt{T_2}}$.
Furthermore at long distance, meron pairs $(-1,0)-(0,-1)$ and $(1,0)-(0,1)$ behave as 
vortices $(-1,-1)$ and $(1,1)$, respectively.

There are classical pictures corresponding to the above eigenstates of 
$(-1,0)-(1,0)$ and $(-1,0)-(0,-1)$ meron pairs.
They are given in Fig.6.
Merons are in the external magnetic field and also have electric charge propotional to
their topological charge.
Therefore, the Lorentz force works for them, as shown by the first and second terms
in (\ref{Lagsol}).
This Lorentz force for merons is balanced by the Coulomb and the confining
linear potentials between merons in a pair.
Therefore,  $(-1,0)-(1,0)$ meron pair moves with a constant velocity and a constant
relative position.
On the other hand in $(-1,0)-(0,-1)$ pair, each meron rotates around the center
of mass coordinate with a constant cyclotron velocity.
These classical pictures support the approximation which ignores
the statistical terms.
For  $(-1,0)-(1,0)$ meron pair, the angle $\theta(\vec{r}_1-\vec{r}_2)$
is a constant of motion in the classical limit and then
${d \over dt}\theta(\vec{r}_1-\vec{r}_2)=0$.
In the path-intergal formalism, paths close to the classical path dominate.
Furthermore, the statistical term is invariant under continuous deformations of path.
Therefore for dominant paths, the statistical term does not contribute
to the path integral in the present case.
We think that this is the reasion why the dispersion relation of vortex pair
in the single-layer case given by (\ref{dis1}) is, at least, qualitatively correct\cite{haldane}.
Similarly for $(-1,0)-(0,-1)$ meron pair, $\theta(\vec{r}_1-\vec{r}_2)\propto t$
in the classical picture, and the statistical term gives only irrelevant effects.
Therefore quantum states of meron pairs which we obtained in the present
discussion are consistent with the approximation which neglects
the statistical terms.
We hope that results of exact diagonalization of the double-layer systems
support out results.

Here however it should be remarked that in the derivation of the Lagrangian of soliton
(\ref{Lagsol}), rotation of solitons etc. have not been taken into account, 
and therefore additional terms like
$(\dot{x}^{\omega,\alpha}_i)^2$, $\dot{x}^{\omega,\alpha}_i
\ddot{x}^{\omega,\alpha}_i$, etc. must appear
as higher-derivative terms in the effective theory  of solitons, if we consider
effects of them.
For exmple, the additional term  in the LLL condition $\epsilon_{ij}\partial_i\hat{J}_j$
induces terms like $(\dot{x}^{\omega,\alpha}_i)^2$.
This situation always happens in deriving an effective theory of topological
objects with finite size, and most of cases low-energy behavior of these objects
is well-described by the low-derivative model like (\ref{Lagsol}), though 
many ideas have been proposed for modification of that.
Especially for constant motion of $(-1,0)-(1,0)$ meron pair, higher-derivative
terms can be legitimately ignored.

%%%%%%%%%%%%%%%%%%%%%%%%%%%%%%%%%%%%%%%%%%%%%%%%%%
\setcounter{equation}{0}

\section{Conclusion}
In this paper, we have discussed topological excitations in the CS gauge theory
with complex boson fields, which describes DL FQHE.
There are two types of topological excitations in the LLL.
The one is the vortices which have a nontrivial $U(1)$ winding number in the 
{\em common} phase factor of both pseudospin up and down electron fields.
They are quite similar to the vortex of quasi-hole  in the single-layer system.
The others are the merons which are pseudospin textures and have a nontrivial
$SU(2)$ winding number.

We first give qualitative arguments about these excitations and then numerical
solutions.
We derived an effective low-energy theory for them by using duality transformation
and the LLL condition.
It is shown that pseudospin textures are described by the $CP^1$ nonlinear
$\sigma$-model with a CS gauge interaction and a Hopf term.
The Hopf term determines statistical parameters of vortices and merons, and  
the result is consistent with that given by the argument through the Aharonov-Bohm effect.
We also consider quantum system of multi-soliton configuration
by retaining only center coordinates of solitons.
Lagrangian of this quantum system is obtained from the effective low-energy
Lagrangian of solitons.
We took into account the effects of inter-layer tunneling, and determined
energy spectrum of meron pairs.
It is quite interesting to observe these excitations by experiment.
Detailed study of dynamical properties of topological excitations is under study
and results will be reported in future publications.

%%%%%%%%%%%%%%%%%%%%%%%%%%%%%%%%%%%%%%%%%%%%%

\newpage

{\Large Figure Captions} 

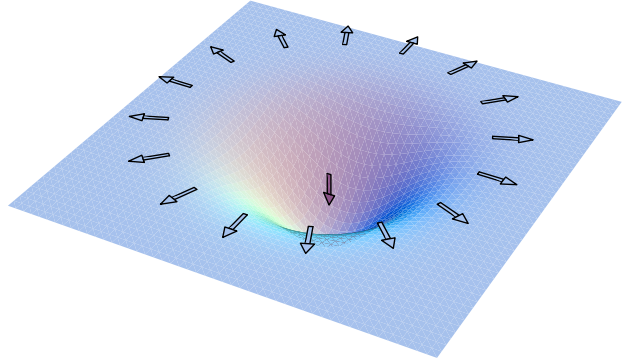
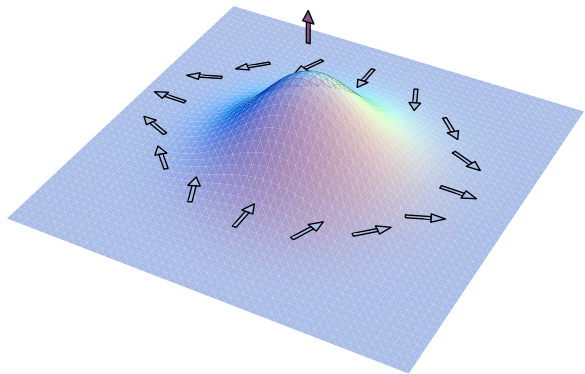
{\bf Fig.1}: Configurations of pseudospin vector of merons. 
Obviously, topological charges of them are fractional.

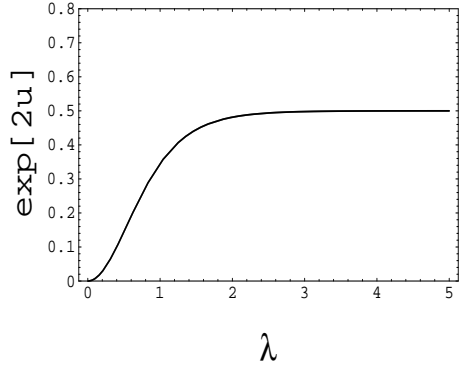
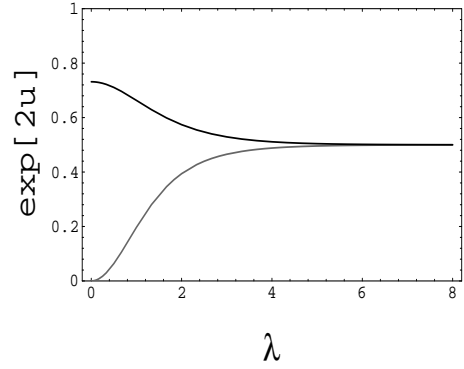
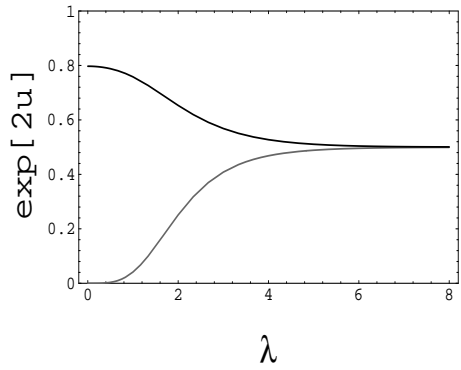
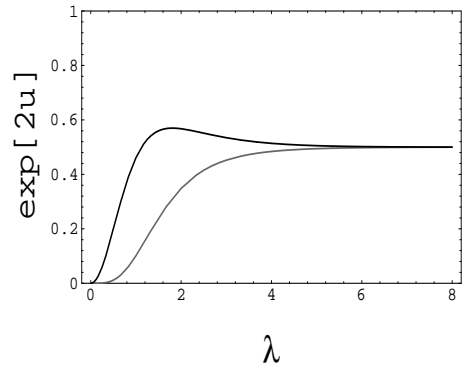
{\bf Fig.2}: Numerical solutions of vortex, merons, etc., in $(3,3,2)$ state of the DL
FQHE.
In both upper and lower layers, charge densities have nontrivial behavior.
This is a result of inter-layer Coulomb interaction.

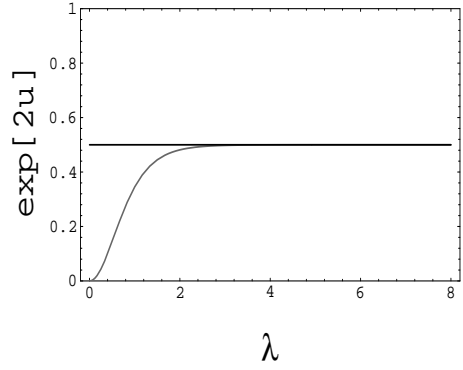
{\bf Fig.3}: Meron solution in the $(5,5,0)$ state. 
There are no correlations between upper and lower layers.

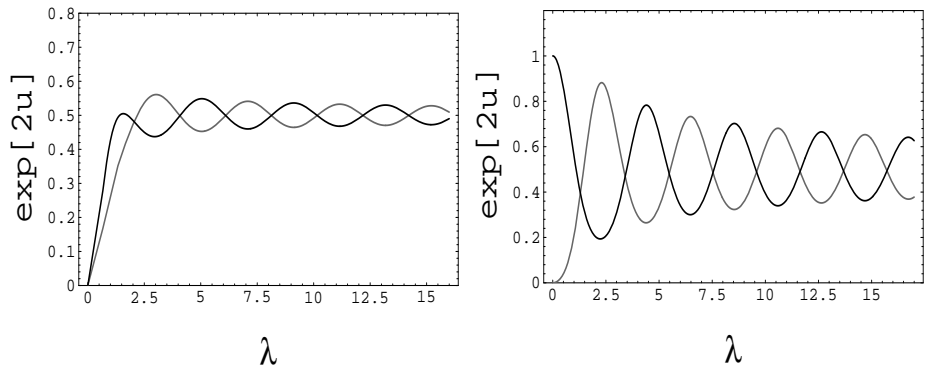
{\bf Fig.4}: Vortex and merons in the $(1,1,4)$ state.
Merons have a long undulating tail because of the dominant inter-layer Coulomb
interaction.

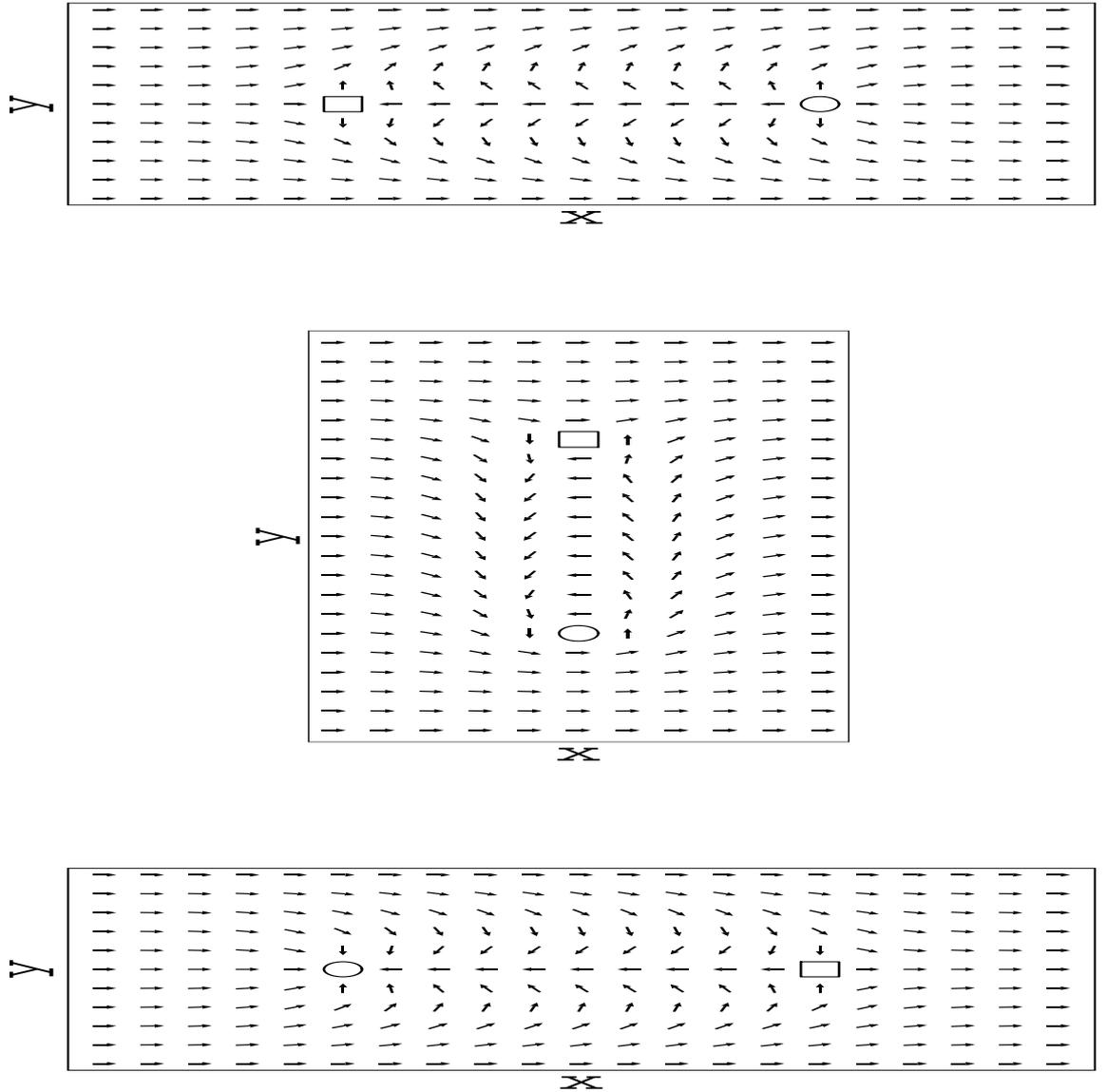
{\bf Fig.5}: Typical configuration of the $(-1,0)-(0,-1)$ meron pair in the presence
of the inter-layer tunneling.
In the region between two merons, direction of pseudospins is reversed, and there
is a confining linear potential between merons.

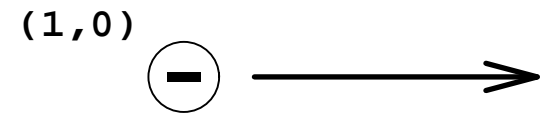
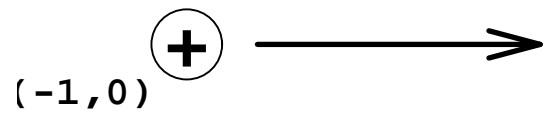
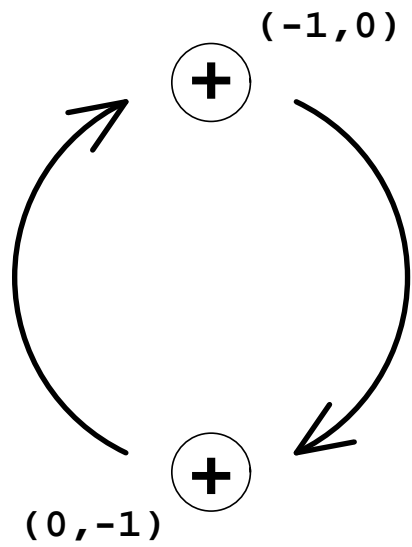
{\bf Fig.6}: Corresponding classical pictures for systems of meron pairs.
Lorentz force is balanced by the Coulomb and the linear interactions.


\newpage
\begin{thebibliography}{1}
\bibitem{sondhi}S.L.Sondhi, A.Karlhede, and S.A.Kivelson, 
Phys.Rev.{\bf B47}(1993)16419;  \\
H.A.Fertig, L.Brey, R.Cote, and A.H.MacDonald,
Phys.Rev.{\bf B50}(1994)11018.
%
\bibitem{exp}S.E.Barrett, G.Dabbagh, L.N.Pfeiffer, K.W.West, and Z.Tycko,  \\
Phys.Rev.Lett.{\bf 74}(1995)5112;  \\
A.Schmeller, J.P.Eisenstein, L.N.Pfeiffer, and K.W.West,  \\
Phys.Rev.Lett.{\bf 75}(1995)4290;  \\
E.H.Aifer, B.B.Goldberg, and D.A.Broido,
Phys.Rev.Lett.{\bf 76}(1996)680.
%
\bibitem{moon}K.Moon, H.Mori, Kun Yang, S.M.Girvin, A.H.MacDonald,
L.Zheng, D.Yoshioka, and S-C.Zhang,
Phys.Rev.{\bf B51}(1995)5138;   \\
Kun Yang, K.Moon, H.Mori, L.Belkhir, S.M.Girvin, A.H.MacDonald, L.Zheng and
D.Yoshioka, 
Phys.Rev.{\bf B54}(1996)11644.
%
\bibitem{CS}S.C.Zhang, H.Hansson, and S.Kivelson,
Phys.Rev.Lett.{\bf 62}(1989)82;  \\
N.Read, 
Phys.Rev.Lett.{\bf 62}(1989)86.
%
\bibitem{Laughlin}R.B.Laughlin,
Phys.Rev.Lett. {\bf 50}(1983)1395.
%
\bibitem{halperin}B.I.Halperin,
Helv.Phys.Acta. {\bf 56}(1983)75.
%
\bibitem{ezawa}Z.F.Ezawa and A.Iwazaki,
Phys.Rev.{\bf B47}(1993)7295.
%
\bibitem{ezawa1}Z.F.Ezawa, M.Hotta, and A.Iwazaki,
Phys.Rev.{\bf B46}(1992)7765.
%
\bibitem{vortex}Z.F.Ezawa, M.Hotta, and A.Iwazaki,
Phys.Rev.{\bf D44}(1991)452.
%
\bibitem{lee}D-H.Lee and C.L.Kane,
Phys.Rev.Lett.{\bf 64}(1990)1313.
%
\bibitem{zhang}M.P.A.Fisher and D.H.Lee,
Phys.Rev.{\bf B39}(1989)2756;  \\
S.C.Zhang,
Int.J.Mod.Phys.{\bf B6}(1992)25.
%
\bibitem{hopf}F.Wilczek and A.Zee,
Phys.Rev.Lett.{\bf 51}(1983)2250.
%
\bibitem{rela}K.Kimm, K.Lee, and T.Lee,
Phys.Rev.{\bf D53}(1996)4436;  \\
Phys.Lett.{\bf B380}(1996)303;  \\
K.Arthur, D.H.Tchrakian, and Yisong Yang,
Phys.Rev.{\bf 54}(1996)5245.
%
\bibitem{IS2}I.Ichinose and A.Sekiguchi, 
paper in preparation.
%
\bibitem{IS3}I.Ichinose and A.Sekiguchi, 
paper in preparation.
%
\bibitem{fetter}A.L.Fetter,
Phys.Rev.{\bf 162}(1967)143;  \\
F.D.M.Haldane and Y.-S.Wu,
Phys.Rev.Lett.{\bf 55}(1985)2887.
%
\bibitem{haldane}F.D.M.Haldane,
in ``{\it The Quantum Hall Effect} " (Springer Verlag, 1990) editated by R.Prange and S.Girvin.
%

\end{thebibliography}
\end{document}